\documentclass[preprint2]{aastex}

\usepackage{graphicx}
\usepackage{graphics}
\usepackage{txfonts}
\usepackage{epsfig}
\usepackage{natbib}
\usepackage{times}
\usepackage{amssymb}
\usepackage{color}

\newcommand\igr{\object{IGR\,J17391$-$3021}}
\newcommand\xtej{\object{XTE\,J1739$-$302}}
\newcommand\ergcms{erg\,cm$^{-2}$\,s$^{-1}$}
\newcommand\ergcmct{erg\,cm$^{-2}$\,ct$^{-1}$}
\newcommand\ergs{erg\,s$^{-1}$}
\newcommand\cmsq{cm$^{-2}$}
\newcommand\integ{{\it{INTEGRAL}}}
\newcommand\suz{{\it{Suzaku}}}
\newcommand\xte{{\it{RXTE}}}
\newcommand\swift{{\it{Swift}}}
\newcommand\asca{{\it{ASCA}}}
\newcommand\xmm{{\it{XMM-Newton}}}
\newcommand\chan{{\it{Chandra}}}
\newcommand\nh{$N_\mathrm{H}$}

\shorttitle{Weak flares from \igr}
\shortauthors{Bodaghee et al.}

\begin{document}

\title{\suz\ observes weak flares from \igr\ \\ representing a common low-activity state in this SFXT}

\author{A. Bodaghee and J. A. Tomsick}
\affil{Space Sciences Laboratory, 7 Gauss Way, University of California, Berkeley, CA 94720, USA}
\email{bodaghee@ssl.berkeley.edu}

\and

\author{J. Rodriguez and S. Chaty}
\affil{Laboratoire AIM, CEA/IRFU - Universit\'e Paris Diderot - CNRS/INSU, \\ CEA DSM/IRFU/SAp, Centre de Saclay, F-91191 Gif-sur-Yvette, France}

\and

\author{K. Pottschmidt}
\affil{CRESST and NASA Goddard Space Flight Center, Astrophysics Science
Division, Code 661, Greenbelt, MD 20771, USA \\
Center for Space Science and Technology, University of Maryland Baltimore
County, 1000 Hilltop Circle, Baltimore, MD 21250, USA}

\and

\author{R. Walter}
\affil{\integ\ Science Data Centre, Universit\'e de Gen\`eve, Chemin d'Ecogia 16, CH--1290 Versoix, Switzerland \\ 
	Observatoire de Gen\`eve, Universit\'e de Gen\`eve, Chemin des Maillettes 51, CH--1290 Sauverny, Switzerland}

\and

\author{P. Romano}
\affil{INAF, Istituto di Astrofisica Spaziale e Fisica Cosmica, \\ Via U. La Malfa 153, I-90146 Palermo, Italy}

\begin{abstract}
We present an analysis of a 37-ks observation of the supergiant fast X-ray transient (SFXT) \igr\ ($=$XTE\,J1739$-$302) gathered with \suz. The source evolved from quiescence to a low-activity level culminating in three weak flares lasting $\sim$3\,ks each in which the peak luminosity is only a factor of 5 times that of the pre-flare luminosity. The minimum observed luminosity was $1.3\times10^{33}$\,\ergs ($d$/2.7\,kpc)$^{2}$ in the 0.5--10\,keV range. The weak flares are accompanied by significant changes in the spectral parameters including a column density (\nh\ $=(4.1_{-0.5}^{+0.4})\times10^{22}$\,cm$^{-2}$) that is $\sim$2--9 times the absorption measured during quiescence. Accretion of obscuring clumps of stellar wind material can explain both the small flares and the increase in \nh. Placing this observation in the context of the recent \swift\ monitoring campaign, we find that weak-flaring episodes, or at least epochs of enhanced activity just above the quiescent level but well below the moderately bright or high-luminosity outbursts, represent more than 60$\pm$5\% of all observations in the 0.5--10\,keV energy range making this the most common state in the emission behavior of \igr.

\end{abstract}


\section{Introduction}

Supergiant Fast X-ray Transients \citep[SFXTs:][]{Smi04,Sgu05,Sgu06,Neg06a} form a new subclass within the group of high-mass X-ray binaries (HMXBs). Previously, HMXBs were catalogued \citep[e.g.][and references therein]{Liu00} into one of either two groups: the transient sources with B-emission line stellar companions (BeXBs); or the persistent, but variable, HMXBs with supergiant O or B-type companions (SGXBs). The SFXTs share properties from both groups: like BeXBs, they are characterized by outbursts of short (hours) to long (days) duration whereby the peak intensity is several orders of magnitude greater than during quiescence; and like SGXBs, they are systems in which a compact object (usually a neutron star) is paired with a supergiant OB star. Their study is therefore important because SFXTs could represent an evolutionary link between BeXBs and SGXBs \citep[e.g.][]{Neg08,Cha10,Bod10,Liu10}.

\object{XTE\,J1739$-$302} \citep{Smi98} was the first hard X-ray transient to display behavior that was identified as characteristic of the class of SFXTs \citep{Neg06a}. It was listed as \object{AX\,J1739.1$-$3020} in the \asca\ catalog of X-ray sources in the Galactic Center region \citep{Sak02}. Five years after its discovery, \integ\ detected a transient source named \igr\ which, because of the positional coincidence and similar emission behavior, was shown to be the hard $X$-ray (18--50\,keV) counterpart to the \xte\ source \citep{Sun03,Smi03}. A refined X-ray position was obtained with \chan\ \citep{Smi06}, which led to the identification of \object{USNO-B1.0\,0596-0585865} ($=$ \object{2MASS\,J17391155$-$3020380}), a supergiant star of spectral type O8\,Iab located $\sim$2.7\,kpc away, as the optical/IR counterpart \citep{Neg06b,Cha08,Rah08}.

This source has been detected numerous times in outburst and in quiescence thanks to monitoring campaigns with \integ\ \citep{Lut05,Sgu05,Tur07,Bla08,Che08,Rom09a,Duc10} and with \swift\ \citep{Rom08a,Rom08b,Sid08,Sid09a,Sid09b,Rom09b,Rom10}. Most of the outbursts are short ($\sim$0.5\,h) and weak ($\sim$10\,cps in 20--40\,keV as seen with \integ), while a few flares last longer than an hour (sometimes days as seen in the 0.5--10\,keV band with \swift) and are rather intense ($\sim$60\,cps or $\sim$400\,mCrab with \integ) \citep{Bla08}. The photoelectric absorption is variable from outburst to outburst \citep{Smi06}, reaching up to $(13^{+4}_{-3})\times10^{22}$\,\cmsq\ during the rising portion of certain flaring episodes \citep{Sid09a,Sid09b}. The dynamic range between quiescence and peak outburst is $\sim10^{4}$ \citep{Int05}. Based on multiple observations of \igr\ gathered over the course of a year with \swift, \citet{Rom09b} showed that bright outbursts account for only 3--5\% of the entire observation set, while the source spends 39$\pm$5\% of the time in an inactive state near quiescence. Quiescence, i.e., a flux in the 0.5--10\,keV range below the \swift\ detection limit of $1.6\times10^{-12}$\,\ergcms\ for a $\sim$1\,ks exposure, is a very rare state \citep{Rom09b}.

What triggers the outbursts is still an open question. Several mechanisms have been proposed to explain the peculiar X-ray emission behavior of SFXTs. The outbursts could be due to one or a combination of the following effects: the intermittent accretion of clumpy wind material by a compact object in a larger orbital radius than in classical SGXBs; passage of the compact object in an eccentric orbit through anisotropic winds emanating from some supergiant stars; and the dynamics of wind accretion onto highly-magnetized neutron stars including gating effects, photo-ionization, transient disks, and Rayleigh-Taylor instabilities \citep[e.g.][and references therein]{Int05,Wal07,Sid07,Neg08,Boz08,Duc09}. 

While the identity of the stellar companion is known, the nature of the compact object has yet to be conclusively determined. The spectral properties of \igr, i.e. its hard power-law shape and possible spectral cutoff around 13\,keV \citep{Sid09a}, suggest a neutron star primary. However, coherent pulsations or cyclotron absorption lines that would firmly establish the presence of a neutron star have not been detected thus far \citep{Smi98,Sgu05,Bla08}. Recently, the orbital period of the system was shown to be 51.47(2)\,d with hard X-ray outbursts ($\gtrsim$20\,keV) and quiescent epochs coexisting at periastron suggesting clumpy, anisotropic winds \citep{Dra10}.

\begin{figure*}[!t] 
\centering
\includegraphics[width=\textwidth,angle=0]{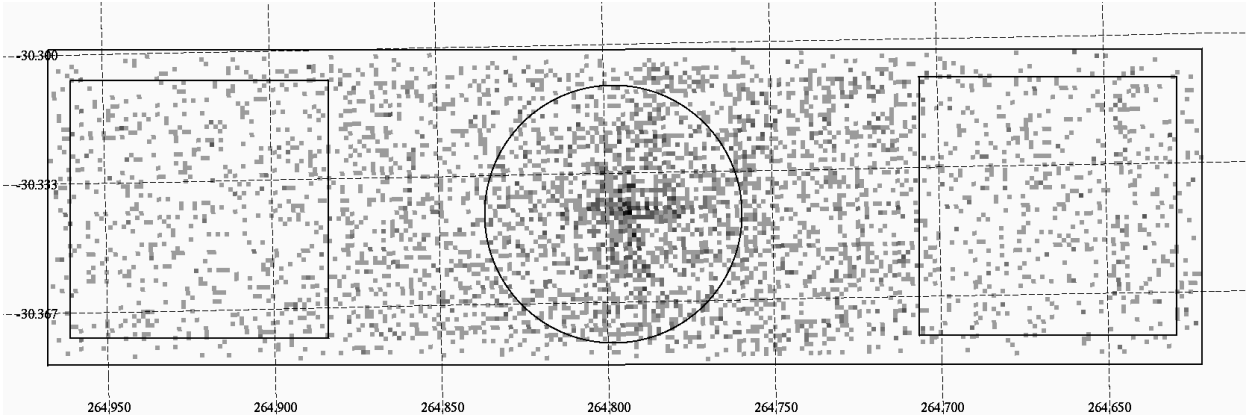}
\caption{Image of IGR\,J17391$-$3021 as captured by the XIS1 detector on {\it Suzaku}. Pixels have been grouped in blocks of 4 (bin: group4 in \texttt{ds9}). The circular source and square background selection regions are also shown. Coordinates are given as equatorial R.A. and Dec. (J2000) where North is up and East is left. }
\label{img}
\end{figure*}

In February, 2008, \suz\ observed \igr\ for 37\,ks of effective exposure time. The data from this observation are described in Section \ref{sec_obs}. Results from timing and spectral analyses are reported in Sections \ref{sec_time} and \ref{sec_spec}, respectively. We discuss our findings and present our conclusions in Sections \ref{sec_disc}--\ref{sec_conc}.

\section{Observations}
\label{sec_obs}

%
\begin{table*}[!t] 
\begin{scriptsize} 
\caption{The different states in the emission behavior of IGR\,J17391$-$3021 as used in the text.}
\vspace{2mm}
\begin{tabular}{ l c c l }
\hline
\hline
state      			& absorbed flux (0.5--10\,keV)				& example time				& reference						\\
				& (\ergcms)							& (MJD)					& 								\\	
\hline
quiescence 		& $\lesssim (1-2)\times10^{-12}$ 			& 54518.495--54518.920	 	& this work, see also \citet{Boz10}						\\	
low 				& $\sim (2-10)\times10^{-12}$	 			& 54518.920--54519.396		& this work, see also \citet{Sid08,Rom09b,Rom10,Boz10}		\\
medium 			& $\sim (2-100)\times10^{-11}$	 			& 54635					& \citet{Rom09b,Rom10}					\\	
high 				& $\gtrsim 10^{-9}$	 					& 54692					& \citet{Rom09b,Rom10}					\\				
\hline
\end{tabular}
\label{tab_states}
\end{scriptsize} 
\end{table*}

The \suz\ space telescope features two co-aligned instruments working in tandem to observe the hard X-ray sky (0.2--600\,keV): the X-ray Imaging Spectrometer \citep[XIS:][]{Koy07} and the Hard X-ray Detector \citep[HXD:][]{Tak07}. Currently, XIS consists of two front-illuminated CCD cameras (XIS0 and XIS3), and one that is back-illuminated (XIS1). 

Observation ID 402066010 (PI: J.A. Tomsick) targeted \igr\ on February 22--23, 2008 (MJD 54518.495--54519.396), for 72.271\,ks. After removing bad events during data reduction (see below), and observation gaps due to the occultation of the satellite by the Earth, the net exposure time that remained was 36.546\,ks. The XIS-nominal pointing mode was used with the quarter-window option in order to gain a higher time resolution. 

Data reduction employed v.6.9 of the HEAsoft software package following the procedures described in the \suz\ ABC Guide v.2. The task \texttt{xispi} converted the unfiltered XIS event files to pulse-invariant channels, which were then input into \texttt{xisrepro} to return cleaned event files.  We extracted the source spectrum from each XIS detector using a circular region of 120$^{\prime\prime}$-radius centered on the position given by \chan\ (see Fig.\,\ref{img}). The background was extracted from a pair of 240$^{\prime\prime}$-wide squares positioned on a source-free region of the detector. Based on these source and background extraction regions, we generated response matrices with \texttt{xisrmfgen} and \texttt{xissimarfgen}.

\begin{figure}[!t] \centering 
\includegraphics[width=8cm,angle=0]{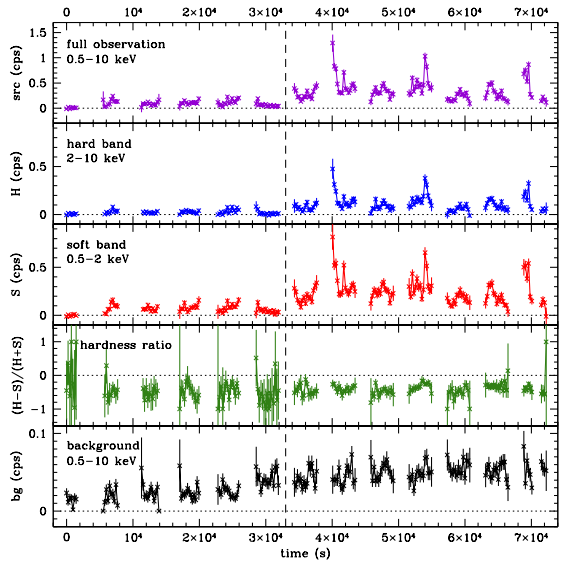}
\vspace{-1cm}
\caption{Light curve of IGR\,J17391$-$3021 from XIS (0.5--10\,keV). The top panel displays the background-subtracted light curve from the source, and the bottom panel shows the background count rate during the observation. Also presented are the source light curves from soft (S) 0.5--2\,keV and hard (H) 2--10\,keV bands, as well as the hardness ratio which is defined as (H$-$S)/(H+S). Each data point collects 240\,s worth of exposure. After 33\,ks of quiescent emission, a low state begins at approximately MJD\,54518.92 (indicated by the vertical dashed line).}
\label{fig_lc}
\end{figure}

To improve photon statistics, we merged the source spectral files for the front-illuminated CCDs (XIS0 and XIS3), and we did the same for the background. Each bin in the XIS spectra contained a minimum of 150 counts. Source and background light curves were generated in 0.5--10\,keV (XIS) with a time resolution of 2\,s. Timing data from the 3 XIS detectors were combined into single light curves for the source and for the background. 

A tuned (v.2.0) background file and the common GTI were used during spectral extraction of HXD-PIN. Dead-time corrections were performed with \texttt{hxddtcor}. The cosmic X-ray background was modeled using the flat-field response file from January 29, 2008, and its contribution was accounted for in the PIN background spectrum. Since \igr\ was very weak during this observation, PIN recorded few high-energy photons. Furthermore, \igr\ is located close to the Galactic Center so diffuse emission from the Galactic Ridge can influence the high-energy spectrum of \igr\ \citep{Yua08}. Applying the best-fitting parameters for the ridge emission from \citet{Val98} to the PIN spectrum, we find that the ridge contributes over 50\% of the PIN photon counts. Hence, the statistics being insufficient for timing and spectral analyses, observations with PIN are not discussed any further.

\begin{figure*}[!t] \centering
\includegraphics[width=\textwidth,angle=0]{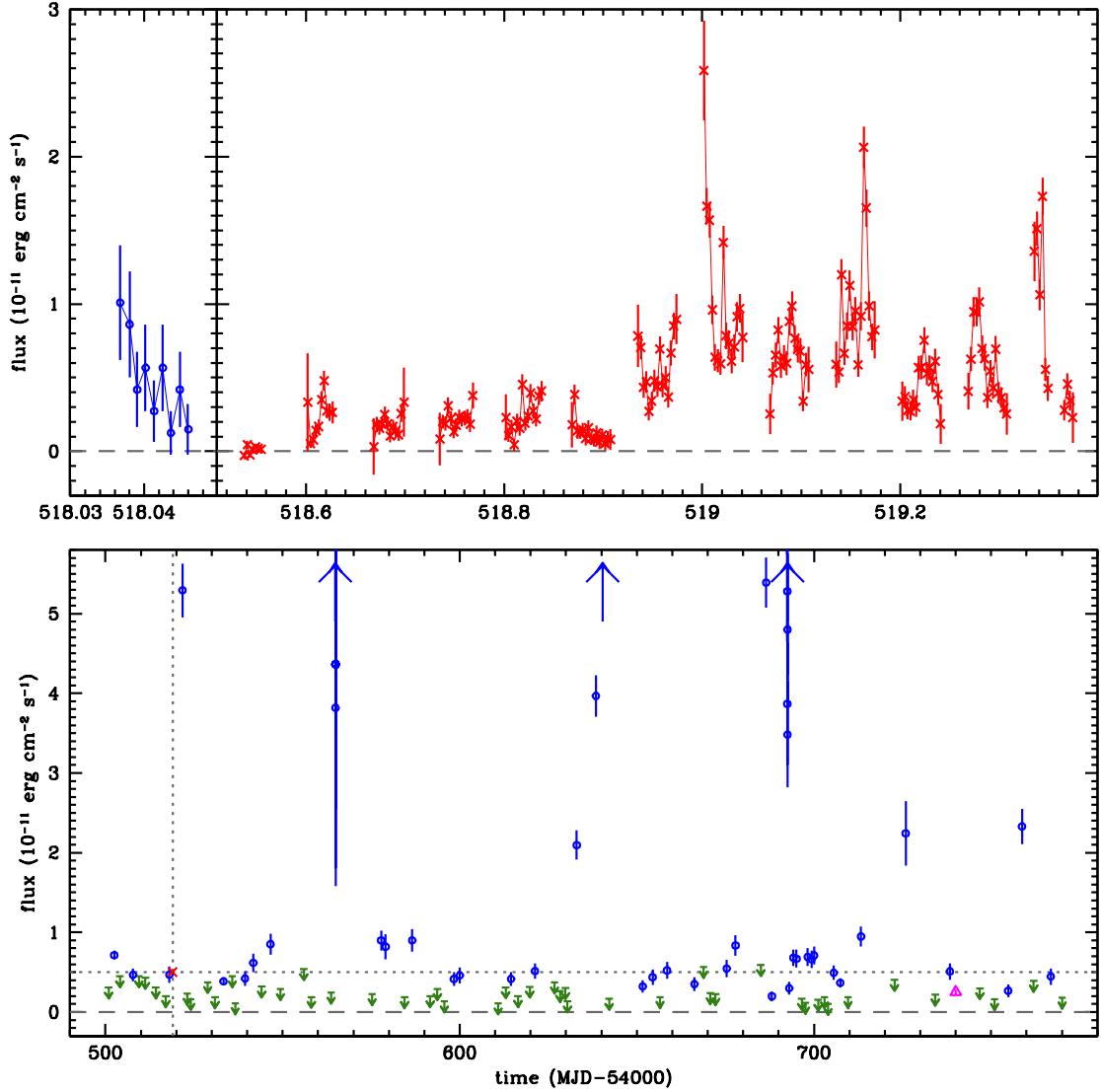}
\vspace{-1cm}
\caption{Light curve of IGR\,J17391$-$3021 from \suz-XIS, \swift-XRT, and \xmm.  Fluxes are in units of $10^{-11}$\,\ergcms\ as observed (i.e. absorbed) in the 0.5--10\,keV energy band, and time is given in MJD ($-$54000). The top panel displays the background-subtracted light curves from the source during Obs.\,ID\,00030987012 from XRT (blue circles, 100-s binning) and from our entire XIS observation (red crosses, 240-s binning). Note the half-day gap between the observations. In the bottom panel, the source light curve based on $\sim$1\,ks observations from the XRT monitoring campaign of \citet{Rom09b} is shown (blue circles) along with their 3$\sigma$ upper limits for non-detections (green downward arrows). The blue upwards arrows designate the maximum flux of high-state outbursts detected by XRT that are situated beyond the scale of the graph (from left to right, their values are 125, 12, and 551). The average flux from our 37-ks XIS observation is plotted as a single red cross at the intersection of the dotted lines. The purple triangle at MJD\,54740 corresponds to the average flux in a 31-ks observation with \xmm\ from \citet{Boz10}. }
\label{fig_lc_all}
\end{figure*}
%

\section{Timing Analysis}
\label{sec_time}

Figure\,\ref{fig_lc} presents the light curve of \igr\ in the 0.5--10 keV band from XIS. In this energy range, the average source count rate (combining XIS 0, 1, and 3) for the entire observation is 0.251$\pm$0.005\,cps. The best-fitting model to the (absorbed) spectrum of the time-averaged \suz\ observation (see Sect.\,\ref{sec_spec}) yields a coefficient of $2.0\times10^{-11}$\,\ergcmct\ which we can use to convert instrumental count rates to units of $10^{-11}$\,\ergcms. Thus, the average flux is $(5.0\pm0.1)\times10^{-12}$\,\ergcms. 

Based on Fig.\,\ref{fig_lc}, we can identify two distinct emission states from \igr\ in this observation. During the initial 33\,ks, \igr\ is in a quiescent state in which the average source count rate is 0.100$\pm$0.009\,cps ($(2.0\pm0.2)\times10^{-12}$\,\ergcms). As with previous observations of \igr\ when it was at or near quiescence \citep{Sid08,Rom09b,Rom10,Boz10}, our \emph{Suzaku} observation shows that even in this state, \igr\ varies on time scales of a few hundred seconds. 

Around 33\,ks after the start of the observation (corresponding to MJD\,54518.92), the source enters a low state in which the mean count rate of 0.358$\pm$0.008\,cps ($(7.2\pm0.2)\times10^{-12}$\,\ergcms) is a factor 3--4 that of the flux in the quiescent state that preceded it. The observations prior to and after MJD\,54518.92 are henceforth referred to in this analysis as ``quiescent'' and ``low'' states, respectively (see Table\,\ref{tab_states}). In addition, three flares lasting $\sim$2\,ks each are revealed in the light curve at $t\sim40$, 54, and 70\,ks. The peak flux of 1.3$\pm$0.2\,cps ($(2.6\pm0.4)\times10^{-11}$\,\ergcms) is an order of magnitude larger than the flux during the quiescent state, but the maxima are still much lower than the typical high-state outbursts expected and seen in such objects. We note that the peaks are nearly equidistant in time being separated by intervals of approximately 15\,ks.

The light curve was divided into two pieces corresponding to a soft (S) band between 0.5 and 2\,keV, and a hard band (H) with energies between 2 and 10\,keV. The hardness ratio can then be defined as (H$-$S)/(H+S) where {a negative value indicates a soft spectrum, and where a positive value denotes a hard spectrum. Figure\,\ref{fig_lc} displays no obvious correlation between the hardness ratio and the source intensity. Binning the light curve into longer time segments, and plotting the hardness ratio versus the source intensity shows only a marginal correlation (but within the errors) between the two parameters at large intensities.

As part of a monitoring campaign focusing on SFXTs, \swift\ targeted \igr\ at various times throughout 2008 \citep{Rom09b}. In particular, there is a $\sim$900-s long XRT pointing (Obs. ID: 00030987012) several hours prior to our \suz\ observation. The 0.5--10\,keV \swift-XRT detections and 3$\sigma$ upper limits are shown along with the average count rate recorded by XIS in Fig.\,\ref{fig_lc_all}. The instrumental conversion coefficient for XRT is $1.3\times10^{-10}$\,\ergcmct\ \citep{Rom09b}.

\section{Spectral Analysis}
\label{sec_spec}

The time-averaged spectrum of \igr\ gathered in the 0.5--10\,keV range by XIS is presented in Fig.\,\ref{fig_spec}. We fit an absorbed power law, employing the abundances of \citet{Wil00} and the photo-ionization cross sections of \citet{Bal92}, to the XIS spectra which yields $\Gamma = $1.4$\pm$0.1 and $N_{\mathrm{H}} = $(3.6$\pm$0.4)$\times10^{22}$\,cm$^{-2}$ (errors are provided at the 90\% confidence level). These are compatible with the values derived by \citet{Sid08} for out-of-outburst epochs in a similar energy range with \emph{Swift}-XRT, with an average luminosity that is slightly higher at $4.8\times10^{33}$\,\ergs\ compared with $3.9\times10^{33}$\,\ergs\ for \citet{Sid08}. Both studies assume a source distance of 2.7\,kpc \citep{Rah08}. The model fits the data well ($\chi_{\nu}^{2} = 0.82$ for 66 degrees of freedom: dof). 

Residuals remain at low energies ($\lesssim$1\,keV) which could hint at the presence of soft excess emission. We know that this source exhibits a soft excess since this emission was detected with \xmm\ \citep{Boz10}. To account for this feature, we added a blackbody whose temperature was fixed to the best-fitting value from \xmm\ ($kT = 1.2$\,keV). This reduces the $\chi_{\nu}^{2}$ to 0.55 for 65 dof ($F$-test probability of 5$\times10^{-5}$), and \nh\ $=(1.71_{-0.26}^{+0.91})\times10^{22}$\,cm$^{-2}$, but the photon index is poorly constrained ($\Gamma = 0.2_{-1.4}^{+0.3}$).

We split the spectra into two segments corresponding to the quiescent and low states described in Sect.\,\ref{sec_time} (i.e. before and after MJD\,54518.92). Table\,\ref{tab_spec}, which lists the spectral parameters of \igr\ for each epoch, shows that there is a clear distinction in column density, photon index, and luminosity. The \nh\ is at least twice as high in the low state ($(4.1_{-0.4}^{+0.5})\times10^{22}$\,cm$^{-2}$) as it is in the quiescent phase ($(1.0_{-0.5}^{+0.6})\times10^{22}$\,cm$^{-2}$). The source spectrum is softer ($\Gamma = 1.5\pm0.1$) during the low state compared to the quiescent state ($\Gamma = 1.0\pm0.3$). The unabsorbed 0.5--10\,keV luminosity rises from 1.3 when the source is quiescent to 7.4$\times10^{33}$\,\ergs\ when it is in the low state. 

Figure\,\ref{fig_con} illustrates the incompatibility (at $>$99\% confidence) in the absorbing column from quiescence to the low state. It is clear from this figure that the column density is the main parameter that evolves from quiescent to low states. Indeed, for a constant $\Gamma \sim 1.3$, the \nh\ ranges from $\sim$1 to $\sim$$4\times10^{22}$\,cm$^{-2}$ between quiescent and low states, respectively. This is different from what was seen by \citet{Boz10} who found that the column density remained steady (within the statistical errors) during the low state while the $\Gamma$ varied by around 50\%. 

To test whether we overestimated the column density during the low state by not considering a soft excess, we included a blackbody component in the spectral fits to quiescent and low states, and we fixed its temperature to the optimal value from \citet{Boz10} (1.2\,keV). The absorbing columns remain inconsistent (at the 90\% confidence level) between quiescence (\nh\ $\le 1.35 \times10^{22}$\,cm$^{-2}$) and the low state (\nh\ $= (1.99_{-0.27}^{+0.79}) \times10^{22}$\,cm$^{-2}$). When the blackbody is fixed to the lower temperature of 0.13\,keV provided by \citet{Boz10}, the column densities are compatible within the large error bars between the quiescent (\nh\ $= (2.5_{-2.0}^{+5.3}) \times10^{22}$\,cm$^{-2}$) and low states (\nh\ $= (5.8_{-1.1}^{+1.4}) \times10^{22}$\,cm$^{-2}$), but in this case, the resulting blackbody radius is $\gtrsim$100\,km which is much larger than the expected size of the compact accretor.

For ease of comparison with previous studies, we fit the source spectrum with models used for this source by other authors \citep{Sid08,Rom09b,Boz10}. An absorbed blackbody provides a better fit than the absorbed power law ($\chi_{\nu}^{2} = 0.71$ for 66 dof). The parameters from this model agree with those describing the blackbody-modeled spectrum of \igr\ when it was in the low state as seen with \swift\ \citep{Sid08,Rom09b}. The results of this model fit to the various states (i.e. time-averaged observation, quiescence, and low state) are listed in Table\,\ref{tab_spec}. Once again, statistically significant differences emerge for the column density (as well as for the radius of the blackbody emission region) for spectra collected in quiescence (\nh\ $\le 0.22 \times10^{22}$\,cm$^{-2}$) and during the low state (\nh\ $= (1.4_{-0.2}^{+0.3})\times10^{22}$\,cm$^{-2}$). 

In the low state, a power law with a partial-covering absorber and an additional component to account for Galactic absorption (fixed to $1.3\times10^{22}$\,cm$^{-2}$ from \citet{Kal05}) yields \nh\ $= (5.1_{-1.5}^{+1.0})\times10^{22}$\,cm$^{-2}$ with a covering fraction of $91\pm4\%$ and $\Gamma =1.76_{-0.25}^{+0.14}$ ($\chi_{\nu}^{2}/$dof $= 0.56/53$). In other words, during the low state that we observed, a large percentage of the X-ray emission is shielded by an absorber whose \nh\ is higher than the expected interstellar value. We could not constrain the value of the partially-absorbing column in the quiescent state.

\begin{figure}[!t] 
\includegraphics[width=5.4cm,angle=-90]{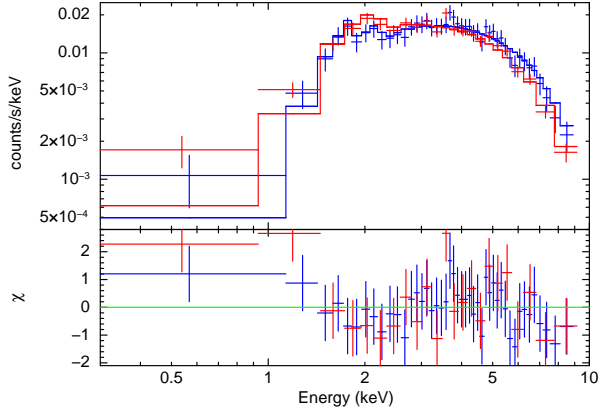}
\caption{Background-corrected spectrum of IGR\,J17391$-$3021 fit with an absorbed power law. The data represent photon counts from XIS1 (red), and XIS0 combined with XIS3 (blue). Each bin collects a minimum of 150 counts.}
\label{fig_spec}
\end{figure}
%

%
\begin{table}[!t]
\begin{scriptsize} 
\caption{Spectral parameters from a power-law (PL), a cutoff power law (CP), and a blackbody (BB) model fit to XIS 0--3 spectra of IGR\,J17391$-$3021. Fluxes ($10^{-12}$\,erg\,cm$^{-2}$\,s$^{-1}$) and luminosities ($10^{33}$\,erg\,s$^{-1}$) are given as unabsorbed in the 0.5--10 keV range for a distance of 2.7\,kpc. Quiescence and the low state refer to pre and post-MJD\,54518.92 spectra, respectively (see Fig.\,\ref{fig_lc}). The \nh\ is given in units of $10^{22}$\,cm$^{-2}$, the cutoff energies and blackbody temperatures are in keV, and the blackbody radii are in meters. Errors are quoted at 90\% confidence.}
\vspace{2mm}
\begin{tabular}{ l c c c c   }
\hline
\hline
model 	& parameter		& time-averaged		& quiescence			& low state			\\
\hline

\noalign{\smallskip}
PL 	& $N_{\mathrm{H}}$		& $3.6\pm0.4$			& $1.0_{-0.5}^{+0.6}$	& $4.1_{-0.4}^{+0.5}$	\\
	& $\Gamma$			& $1.4\pm0.1$			& $1.0\pm0.3$			& $1.5\pm0.1$			\\
	& $F$				& 5.5					& 1.5					& 8.5					\\
	& $L$				& 4.8					& 1.3					& 7.4					\\
	& $\chi_{\nu}^{2}$/dof	& 0.82/66				& 0.68/8				& 0.73/54				\\
\noalign{\smallskip}
\hline

\noalign{\smallskip}
CP 	& $N_{\mathrm{H}}$		& $2.1\pm0.3$			& $\le 1.5$			& $2.2_{-1.2}^{+0.7}$	\\
	& $\Gamma$			& $-0.2_{-0.8}^{+0.6}$	& $-0.1\pm1.3$			& $-0.3_{-1.0}^{+0.6}$	\\
	& $E_{\mathrm{cut}}$	& $3_{-1}^{+2}$			& $\ge 2$				& $3_{-1}^{+1}$			\\
	& $F$				& 4.0					& 1.3					& 5.7					\\
	& $L$				& 3.5					& 1.1					& 4.9					\\
	& $\chi_{\nu}^{2}$/dof	& 0.57/65				& 0.64/7				& 0.44/53				\\
\noalign{\smallskip}
\hline

\noalign{\smallskip}
BB 	& $N_{\mathrm{H}}$		& $1.2\pm0.2$			& $\le 0.22$			& $1.4_{-0.2}^{+0.3}$	\\
	& $kT_{\mathrm{bb}}$	& $1.49_{-0.05}^{+0.06}$	& $1.57_{-0.15}^{+0.16}$	& $1.49_{-0.06}^{+0.07}$	\\
	& $R_{\mathrm{bb}}$	& $72\pm5$			& $37_{-4}^{+8}$		& $88_{-6}^{+7}$		\\
	& $F$				& 3.4					& 1.2					& 5.1					\\
	& $L$				& 3.0					& 1.0					& 4.4					\\
	& $\chi_{\nu}^{2}$/dof	& 0.71/66				& 0.73/8				& 0.51/54				\\
\noalign{\smallskip}
\hline

\end{tabular}
\label{tab_spec}
\end{scriptsize}
\end{table}

%
\begin{table}[!t]
\begin{scriptsize} 
\caption{Same as Table\,\ref{tab_spec}, but for key parameters from rate-selected spectra from epochs of low count rate (LCR), medium count rate (MCR), and high count rate (HCR). }
\vspace{2mm}
\begin{tabular}{ l c c c c   }
\hline
\hline
model 	& parameter		& LCR				& MCR				& HCR				\\
\hline

\hline

\noalign{\smallskip}
	& count rate range (cps)	& $<$0.1				& 0.1--0.3				& $>$0.3				\\ 
	& average count rate	 (cps) & 0.05				& 0.19				& 0.45				\\
	& exposure time (ks)		& 22					& 11					& 4					\\
\noalign{\smallskip}
\hline

\noalign{\smallskip}
PL 	& $N_{\mathrm{H}}$		& $\le 0.9$			& $4.1_{-0.9}^{+1.0}$	& $4.3_{-0.8}^{+1.0}$	\\
	& $\Gamma$			& $0.6_{-0.3}^{+0.2}$	& $1.7\pm0.3$			& $1.4\pm0.2$			\\
	& $\chi_{\nu}^{2}$/dof	& 1.2/11				& 0.83/34				& 0.50/38				\\
\noalign{\smallskip}
\hline

\noalign{\smallskip}
CP 	& $N_{\mathrm{H}}$		& $\le 0.2$			& $2.1_{-0.5}^{+1.2}$	& $2.5_{-0.6}^{+0.8}$	\\
	& $\Gamma$			& $-0.8\pm0.2$			& $-0.1_{-0.2}^{+0.4}$	& $-0.3\pm0.2$			\\
	& $E_{\mathrm{cut}}$	& $3$ (fixed)			& $3$ (fixed)			& $3$ (fixed)			\\
	& $\chi_{\nu}^{2}$/dof	& 1.2/10				& 0.71/34				& 0.27/38				\\
\noalign{\smallskip}
\hline

\noalign{\smallskip}
BB 	& $N_{\mathrm{H}}$		& $\le 0.1$			& $1.2_{-0.4}^{+0.5}$	& $1.2_{-0.4}^{+0.5}$	\\
	& $kT_{\mathrm{bb}}$	& $1.8_{-0.2}^{+0.3}$	& $1.4\pm0.1$			& $1.4\pm0.1$			\\
	& $\chi_{\nu}^{2}$/dof	& 1.6/10				& 0.72/34				& 0.72/34				\\
\noalign{\smallskip}
\hline

\end{tabular}
\label{tab_spec2}
\end{scriptsize}
\end{table}

A cutoff power law model gives a good fit to the time-averaged spectrum ($\chi_{\nu}^{2}/$dof $= 0.57/65$). The power law index is $\Gamma=-0.2_{-0.8}^{+0.6}$ with a cutoff at $E_{\mathrm{cut}}=3_{-1}^{+2}$\,keV, and the column density is \nh\ $= (2.1\pm0.3)\times10^{22}$\,cm$^{-2}$. These are consistent with the parameters listed in \citet{Boz10}. The index and cutoff energy are compatible between the quiescent ($\Gamma=-0.1_{-1.3}^{+1.3}$, $E_{\mathrm{cut}}>2$\,keV) and low ($\Gamma=-0.3_{-1.0}^{+0.6}$, $E_{\mathrm{cut}}=3_{-1}^{+1}$\,keV) states. In this case, the upper limit to the column density during quiescence ($< 1.5\times10^{22}$\,cm$^{-2}$) lies within the 90\% confidence interval of \nh\ in the low state ($(2.2_{-1.2}^{+0.7})\times10^{22}$\,cm$^{-2}$).

The spectra were also fit with the Mewe-Kaastra-Liedahl (MEKAL) model employing a dual absorber: one that affects the system's soft X-ray emission from hot diffuse gas in the supergiant star's wind (\nh$_{\mathrm{sg}}$); and another that impacts the high-energy power-law emission from the compact accretor (\nh$_{\mathrm{x}}$). The gas temperature was set to 0.15\,keV \citep{Boz10}. In the quiescent state, we obtain \nh$_{\mathrm{sg}} \le 2.2\times10^{22}$\,cm$^{-2}$ and \nh$_{\mathrm{x}} \le 2.4\times10^{22}$\,cm$^{-2}$ for $\Gamma$ fixed to the optimal value of 1.0 from the absorbed power law fit to the quiescent spectrum ($\chi_{\nu}^{2}/$dof $= 0.54/52$). For the spectrum from the low state, \nh$_{\mathrm{sg}} = (5.8_{-2.0}^{+0.7})\times10^{22}$\,cm$^{-2}$ and \nh$_{\mathrm{x}} \le 1.4\times10^{22}$\,cm$^{-2}$ for $\Gamma = 1.8\pm0.2$ ($\chi_{\nu}^{2}/$dof $= 0.55/65$). This suggests that not only is there material around the binary system absorbing X-rays, there is additional matter located near the compact object.

\begin{figure}[!t] 
\includegraphics[width=8cm,angle=0]{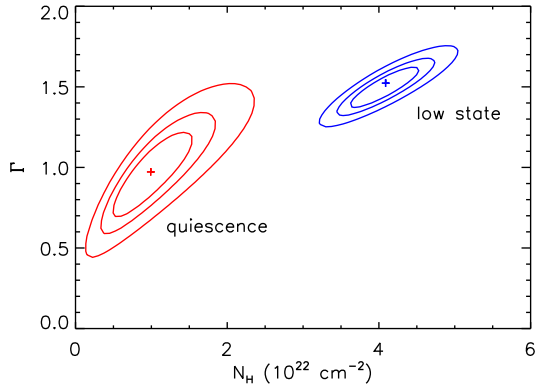}
\vspace{-8mm}
\caption{Parameter space of \nh\ and $\Gamma$ derived from absorbed power laws fit to the spectra of IGR\,J17391$-$3021 during quiescent and low states. The best-fitting parameters are represented by crosses and the contours denote confidence levels of 68\%, 90\%, and 99\%.}
\label{fig_con}
\end{figure}

We sought to determine how the spectral parameters varied with the source intensity. So we created spectra from time intervals in which the count rates (CR) were $<$0.1\,cps (``low CR'' or LCR), between 0.1 and 0.3\,cps (``medium CR'' or MCR), and $>$0.3\,cps (``high CR'' or HCR). Then, each rate-resolved spectrum was fit with an absorbed power law, an absorbed cutoff power law with $E_{\mathrm{cut}}$ fixed to our best-fitting value (3\,keV), and an absorbed blackbody model. Table\,\ref{tab_spec2} lists the results of the spectral fits. Regardless of the spectral model, the \nh\ is significantly smaller during the LCR epochs. The MCR and HCR epochs have spectra that are consistent, and both are statistically incompatible with the spectrum from LCR epochs.

\section{Discussion}
\label{sec_disc}

\subsection{Weak Flares and Low Activity}

Our analysis of a \suz\ observation of \igr\ revealed the source evolving from quiescence to a low-activity state. Yet even at this low-flux level, the emission from \igr\ is characterized by variability on time scales of hundreds to thousands of seconds indicating an accretion that is continuous and dynamic. The average luminosity of \igr\ during the entire \suz\ observation is $4.8\times10^{33}$\,\ergs\ (unabsorbed) and $3.1\times10^{33}$\,\ergs\ (absorbed) in the 0.5--10\,keV range, and assuming a source distance of 2.7\,kpc \citep{Rah08}. 

During the quiescent state, which consists of the first $\sim$33\,ks of the observation, the luminosity in the 0.5--10\,keV band is only $1.1\times10^{33}$\,\ergs\ as observed, and $1.3\times10^{33}$\,\ergs\ when corrected for absorption. This is equivalent to the lowest luminosity detected by \asca\ \citep{Sak02}. In our flux-limited spectrum (LCR), the unabsorbed luminosity is $8.7\times10^{32}$\,\ergs, i.e., twice the minimum values measured with \xmm\ \citep{Boz10} and \swift\ \citep{Rom10}.

Concluding our observation is a $\sim$39-ks epoch of enhanced activity including three flares lasting 1--3\,ks each where the peak luminosities are only a factor of $\sim$5 greater than the quiescent baseline emission that preceded it. Clearly, these flares are significantly weaker than the typical high-luminosity outbursts of SFXTs in which the dynamic range is a few orders of magnitude. These small spikes are reminiscent of the short, low-intensity flares seen in this source by \citet{Boz10} with \xmm. 

Separated by regular intervals of approximately 15\,ks, the flares appear to be quasi-periodic. This timescale is well short of the established orbital period of 51\,days \citep{Dra10}, and it is probably too long for a pulsation period if the compact object is, indeed, a neutron star. \citet{Duc10} detected quasi-periodic outbursts of similar frequency from this source in \integ\ observations, but for longer flares (between 5 and 10\,ks) of much higher intensity ($L_{\mathrm{x}} \sim10^{36}$\,\ergs). They ascribe the quasi-periodic flaring to the formation and dissipation of an unstable accretion disk which is the mechanism favored by \citet{Smi06} to explain bright outbursts in this source. The pulsation period of the suspected neutron star has never been measured, despite numerous observations in and out of outburst. If the pulse period of the suspected NS in this system were known, then a change in the sign of the period derivative could indicate a dissipation and reformation of a disk \citep{Taa88}. This 15-ks frequency could possibly correspond to a periodicity but we would require further observations to confirm or reject this possibility.

\subsection{Spectral Evolution}

The evolution of the emission from quiescence to the low state is accompanied by changes in the X-ray spectral parameters. The \nh\ is higher and the spectral slope is steeper during the low state than during quiescence. Similarly, the high count-rate resolved spectrum showed larger \nh\ and photon index than the low count-rate resolved spectrum. The main parameter that evolves is the column density. This is different from the weak flares seen by \citet{Boz10} who found that as the average count rate of the spectrum increased, $\Gamma$ became smaller (i.e. flatter, harder spectra, and more positive hardness ratios) while the \nh\ did not change significantly.

Differences in the spectral properties of the flares caught by \suz\ and the ones seen with \xmm\ could be due to unequal geometric configurations of the accretion stream between the observations. The capture by the NS of a clump of wind material will lead to an increase in the mass-accretion rate which both instruments detect as a spike in the X-ray luminosity. A partial-covering model fit to the X-ray spectra from the weak flares observed by \citet{Boz10} suggests that up to 40\% of the X-ray emission is not absorbed by the clump itself, but rather by the interstellar medium. The same model fit to our low-state spectrum reveals an absorber that shields a much larger fraction of the X-ray source (less than 13\% of the emission is unblocked).

On the other hand, if prior to being accreted, the captured material crosses the line of sight between the observer and the NS, the flare will be accompanied by an increase in the column density, which is precisely what we observe \citep[see also][]{Ram09}. The expected interstellar value along the line of sight is (1.2--1.3)$\times10^{22}$\,cm$^{-2}$ \citep{Dic90,Kal05}, consistent with what we measured during quiescence. This represents an upper limit in the direction of the source since it results from the integration of absorbing material through the entire Galaxy, while \igr\ is located only $\sim$3\,kpc away. The \nh\ from a power law fit to the spectrum of \igr\ during its low state is 2--4 times the interstellar value and 2--9 times the quiescent value. This suggests that the source is not strongly absorbed (intrinsically) during this observation except when the NS accretes a clump passing along our line of sight. Material in a clump located near the NS will be photo-ionized so the true column density is expected to be higher than our measured value which assumes cold neutral gas. Our largest column density remains an order of magnitude below the maximum \nh\ measured for this source, $(32\pm3)\times10^{22}$\,cm$^{-2}$ \citep{Smi06}, although their result relied on a spectrum from \xte\ that did not extend below 2.5\,keV.

If the origin of the hard X-ray emission is a blackbody, then this radius expands from $37_{-4}^{+8}$\,m during quiescence to $88_{-6}^{+7}$\,m when the source is in the low state, in agreement with \citet{Rom09b} who showed the blackbody radius growing with intensity. It is tempting to associate this small blackbody-emitting region with the polar caps of a magnetized accreting neutron star \citep[$R_{\mathrm{bb}} \lesssim 0.1 R_{\mathrm{NS}}$:][]{Hic04}. However, \citet{Boz10} caution that models describing spherically-symmetric accretion onto magnetized neutron stars predict an inverse relation between the size of the hot spot and the square root of the X-ray luminosity, at least for systems with an average $L_{\mathrm{X}} < 10^{35}$\,\ergs\ \citep[][and references therein]{Whi83}. 

\subsection{The Ubiquity of Low Activity}

Our \suz\ observation of \igr, when placed in the context of the long-term monitoring campaign of this source by \swift-XRT \citep{Rom09b}, shows that the average flux that we recorded is consistent with many of the XRT detections at a low flux level ($F \sim (2-10)\times10^{-12}$\,\ergcms), and is on par with a few of the 3$\sigma$ upper limits to the count rates of non-detections as measured by XRT (Fig.\,\ref{fig_lc_all}). 

Weak flares that are just above quiescence, and whose peak fluxes are at the level of the faint detections in the \swift\ monitoring data, compels us to propose four states in the emission behavior of \igr: the extremely low but variable emission of quiescence (average $F_{\mathrm{abs}} \lesssim (1-2)\times10^{-12}$\,\ergcms\ in 0.5--10\,keV); a low-activity state that potentially includes weak flares such as those seen here and in \citet{Boz10} ($F \sim (2-10)\times10^{-12}$\,\ergcms); a medium state with flares of larger amplitude ($F \sim (2-10)\times10^{-11}$\,\ergcms); and finally, a high state in which the outbursts represent the maximum source luminosity ($F \gtrsim 10^{-9}$\,\ergcms). The boundaries are not strict since the peak fluxes in the low state extend up to $3\times10^{-11}$\,\ergcms, but they provide a useful classification for the various emission states in this source.

If we assume that the \swift\ detections when the count rates were sufficiently low (i.e. $F \sim (2-10)\times10^{-12}$\,\ergcms) were due to low states that may or may not include weak flares, then we can define a low-state duty cycle (DC) as follows:

\begin{equation}
\mathrm{DC} = \frac{ \Sigma T_{\mathrm{low}} } { \Sigma T_{\mathrm{tot}} }  = 60\pm5\%
\end{equation}

Here, $\Sigma T_{\mathrm{low}}$ is the sum of exposure times of all observations in which \igr\ is detected at an average flux within the range of $F \sim (2-10)\times10^{-12}$\,\ergcms, including the \suz\ and \xmm\ observations, while $\Sigma T_{\mathrm{tot}}$ denotes the total exposure from all observations. As expected, the duty cycle is not far from the value of 61$\pm$5\% provided in \citet{Rom09b,Rom10}. 

Low-activity states in which the flux is above the ``typical" quiescent level, but well below the medium and bright flaring episodes, are therefore quite common in this source. In the field of \igr, the limiting flux for XRT is $1.6\times10^{-12}$ and $1.5\times10^{-13}$\,\ergcms\ for exposure times of 1 and 30\,ks, respectively \citep{Rom09b}. Therefore, most \swift\ upper limits would be detections if the exposures had been equivalent to those of the \suz\ and \xmm\ observations suggesting that the duty cycle of the low state is much higher. 

According to the source ephemeris \citep{Dra10}, the midpoint of our observation coincides with an orbital phase of $0.37_{-0.01}^{+0.02}$ whereas the \xmm\ observation took place at phase 0.8. This is another indication that the low state is common since it is not confined to a specific part of the orbit.

Despite the large duty cycle, weak flares lasting a few hundred to a few thousand seconds can be missed by short monitoring observations. Because their peak flux is so low, detecting a low state requires prolonged observations where the long integration time guarantees that we will achieve the sensitivity necessary to detect such low levels of activity. Given that \igr\ is a prototypical member of its class, it would not be surprising to find weak flares from other SFXTs as telescopes continue to devote observing time to such objects. In fact, recent \suz\ and \xmm\ observations of the SFXT \object{IGR\,J08408$-$4503} included similar low states that were punctuated by weak flares \citep{Boz10,Sid10}.

\section{Summary \& Conclusions}
\label{sec_conc}

The main results from our \suz\ observation of the SFXT \igr\ can be summarized as follows:

\begin{itemize}

\item[-]For most of the observation, \igr\ is in what we believe represents the quiescent state. The unabsorbed luminosity in the 0.5--10\,keV range is $1.3\times10^{33}$\,\ergs, and as low as $8.7\times10^{32}$\,\ergs\ in the flux-limited spectrum. The spectrum from this epoch is well described by an absorbed, hard power law with \nh\ $=(1.0_{-0.5}^{+0.6})\times10^{22}$\,cm$^{-2}$ and $\Gamma = 1.0\pm0.3$, or by an absorbed thermal blackbody with a temperature of $1.57_{-0.15}^{+0.16}$\,keV. 

\item[-]Around 33\,ks into the observation (MJD\,54518.92) the source became slightly more active (``low state'') culminating in a series of weak flares each lasting $\sim$3\,ks whose peak luminosity is only a factor of $\sim$5 greater than the quiescent emission that preceded it. During this low state, the \nh\ reached $(4.1_{-0.4}^{+0.5})\times10^{22}$\,cm$^{-2}$ which is a factor of $\sim$2--4 times that of the expected interstellar value, and which is $\sim$2--9 times the \nh\ measured during quiescence. The accretion of small clumps of stellar wind material can be responsible for the weak flares, while the increase in \nh\ can be explained by the passage of the clump in the line-of-sight to the X-ray source prior to accretion. 

\item[-]The average luminosity of the entire observation is $4.8\times10^{33}$\,\ergs. When placed in the full light curve from the \swift-XRT monitoring campaign by \citet{Rom09b}, we find that this luminosity is on par with the low-intensity XRT detections (as well as a few of the 3$\sigma$ upper limits from non-detections). Furthermore, we suspect that many of the faintest detections from the XRT contain similar low states (potentially including weak flares) that are just above the quiescent flux. We estimate that the duty cycle of the low state is 60$\pm$5\%, which makes it the most common state for this source.

\end{itemize}

As a prototypical member of the class of SFXTs, \igr\ holds valuable clues to the accretion processes of these intriguing transients. Continued monitoring of this source by sensitive instruments such as \suz\ should fuel the ongoing debate concerning the mechanisms that are responsible for their peculiar emission behavior. 

\acknowledgments
The authors thank the anonymous referee whose revision of the draft led to significant improvements in the quality of the manuscript. AB and JT acknowledge \suz\ Guest Observer Grant NNX08AB88G and \chan\ Grant G089055X. This research has made use of: data obtained from the High Energy Astrophysics Science Archive Research Center (HEASARC) provided by NASA's Goddard Space Flight Center; the SIMBAD database operated at CDS, Strasbourg, France; NASA's Astrophysics Data System Bibliographic Services.

{\it Facilities:} \facility{Suzaku}, \facility{Swift}, \facility{XMM-Newton}.

\bibliographystyle{apj}
\bibliography{bodaghee_2010_02.bib}
\clearpage

\end{document}